\documentclass[9pt,twocolumn,twoside]{osajnl}
\usepackage{eurosym}
\usepackage{exscale}
\usepackage{siunitx}
\usepackage{relsize}
\usepackage[sort&compress]{natbib}
\usepackage{graphicx,subfigure,amssymb,dcolumn,bm,amsmath}
\journal{ol}
\setboolean{shortarticle}{true}

\title{Stable high-order solitons in spiral potentials}
\author[1*]{Liangwei Dong}
\author[1]{Jingyi Deng}
\author[2]{Changming Huang}
\author[3]{Boris A. Malomed}
\affil[1]{Department of Physics, Zhejiang University of Science and Technology, Hangzhou, 310023, China }
\affil[2]{Department of Physics, Changzhi University, Changzhi, 046011, China}
\affil[3]{Instituto de Alta Investigaci\'{o}n, Universidad de Tarapac\'{a}, Casilla 7D, Arica, Chile}
\affil[*]{Corresponding author: dlw\_0@163.com}

\ociscodes {(190.0190) Nonlinear optics; (190.6135) Spatial solitons; (070.7345) Wave propagation.}
\begin{abstract}
We present a comprehensive study of optical solitons supported by spiral potentials in media with the cubic-quintic (CQ) nonlinearity. A  variety of families of stationary states, including fundamental and high-order (excited) in-phase, out-of-phase, and hybrid-phase ones, are found. The linear stability analysis, corroborated by direct simulations, demonstrates that all upper-branch nonlinear states in potentials with different azimuthal indices are \emph{completely stable}, which rarely occurs in soliton physics. Our findings suggest spiral potentials as an effective means for multistable optical trapping, with potential applications in all-optical data processing.
\end{abstract}
\setboolean{displaycopyright}{true}

\begin{document}
\maketitle

Spatial optical solitons are self-trapped beams with the transverse profile
maintained by the balance of diffraction and nonlinearity \cite%
{kivshar2003optical}. They are well known in a variety of settings, such as
photorefractive crystals \cite{Book2021}, semiconductor waveguides \cite%
{PhysRevLett.134.023802}, liquid crystals \cite{NCLiquidcrystals}, atomic
vapors \cite{NCedgeSolitons}, etc. The design of the spatial
refractive-index patterns, \emph{viz.}, the creation of effective optical
potentials with tailored symmetries, is a core pursuit in nonlinear
photonics \cite{Science20202}. In this context, photonic lattices have
enabled the realization of solitons of the gap, surface, and discrete-vortex
types \cite{Chen_2012}. More complex potentials with nontrivial structures,
such as honeycomb lattices \cite{Mukherjee:23}, quasicrystals \cite%
{YeNatureP2024}, and moir\'{e} patterns \cite{YeNatureP}, have further
expanded the possibilities for controlling the light propagation in
engineered settings.


Among various confining configurations, spiral-shaped potentials are of
particular interest. Spiral structures arise naturally in various physical
settings, from galactic arms \cite{galaxy} to turbulent vortices \cite{turb}%
, chemical waves \cite{Winfree}, and excitations in cardiac tissues \cite%
{Pertsov}. The spirals break continuous translational and rotational
symmetries, while maintaining discrete rotational invariance. The interplay
of their engineered counterparts in photonics with material nonlinearity
suggests possibilities to observe novel guided optical states, as we
demonstrate in this Letter.

In two-dimensional geometries, the Kerr self-focusing induces the
catastrophic wave collapse \cite{malomed2022}. The collapse can, however, be
suppressed by competing nonlinearities \cite%
{Quiroga-Teixeiro97,Pego,michinel2004square}, which occur in diverse
physical contexts, including plasma physics \cite{zakharov1971behavior},
Bose superfluids \cite{PhysRevLett.78.1215}, and quantum droplets \cite%
{dong2021rotating}. In particular, the dielectric response of some materials
used in nonlinear optics is accurately approximated by the cubic-quintic
combination of nonlinear terms \cite{Lawrence:98,Cid}. The CQ model is known
to support stable multipole and vortex solitons \cite%
{Quiroga-Teixeiro97,Pego,michinel2004square,dong2023multipole, Dong:23}.


Here, we report new families of spatial soliton created by the interplay of
spiral potentials with the CQ nonlinearity. Specifically, we find that all
upper-branch solitons (in terms of the dependence of the power on the
propagation constant) are stable in their entire existence domains.

The propagation of optical beams along the $z$-axis in a bulk medium with
the CQ nonlinearity and external potential is governed by the scaled
nonlinear Schr\"{o}dinger equation
\begin{equation}
i\frac{\partial \Psi }{\partial z}=-\frac{1}{2}\nabla _{\perp }^{2}\Psi
-V_{0}V(r,\theta )\Psi -|\Psi |^{2}\Psi +|\Psi |^{4}\Psi .  \label{Eq1}
\end{equation}%
Here $\Psi (x,y,z)$ is the complex field amplitude, $\nabla _{\perp
}^{2}=\partial _{x}^{2}+\partial _{y}^{2}$ is the diffraction operator, $%
(x,y)$ are the transverse coordinates, scaled by the input beam width, $z$
is the propagation distance scaled by the diffraction length, and $V_{0}$ is
the depth of the spiral potential, which is defined as
\begin{equation}
V(r,\theta )=\frac{r}{{r^{2}+r_{0}^{2}}}\cos \left( \frac{2\pi r}{T}-N\theta
\right) ,  \label{Eq2}
\end{equation}%
written in terms of the polar coordinates $\left( r,\theta \right) $, $T$
defines the radial period, and integer $N$ is the azimuthal index
representing the number of spiral arms. Such refractive index modulation can be realized by direct laser writing. 
The potential's amplitude has a maximum at $r=r_{0}$, subsiding along the arms. Equation (\ref{Eq1})
conserves the power (norm), $U=\iint |\Psi (x,y)|^{2}dxdy$, and Hamiltonian,
$H=\iint \left[ \frac{1}{2}\left\vert \nabla \Psi \right\vert
^{2}-V_{0}V(r,\theta )\left\vert \Psi \right\vert ^{2}-\frac{1}{2}\left\vert
\Psi \right\vert ^{4}+\frac{1}{3}\left\vert \Psi \right\vert ^{6}\right]
dxdy $.


Stationary soliton solutions are sought for as $\Psi (x,y,z)=\psi
(x,y)\exp(ibz)$, with real propagation constant $b$ and real stationary
solution $\psi (x,y)$ satisfying the equation
\begin{equation}
\frac{1}{2}\nabla _{\perp }^{2}\psi -b\psi +V_{0}V(r,\theta )\psi +\psi
^{3}-\psi ^{5}=0.  \label{Eq3}
\end{equation}%
The stability of stationary states against small perturbations was checked
by taking a perturbed solution to Eq.~(\ref{Eq1}) as $\Psi (x,y,z)=\left[
\psi (x,y)+u(x,y)\exp (\lambda z)\right. \left. +v^{\ast }(x,y)\exp (\lambda
^{\ast }z)\right] \exp (ibz)$, where $u$ and $v$ are infinitesimal
perturbations, $^{\ast }$ stands for the complex conjugate, and $\lambda $
is the (generally, complex) instability growth rate. The linearization of
Eq.~(\ref{Eq1}) around $\psi $ yields an eigenvalue problem for $u$ and $v$:
\begin{equation}
i%
\begin{bmatrix}
\mathcal{L}_{\text{diag}} & \mathcal{L}_{\text{off}} \\
-\mathcal{L}_{\text{off}}^{\ast } & -\mathcal{L}_{\text{diag}}^{\ast }%
\end{bmatrix}%
\begin{bmatrix}
u \\
v%
\end{bmatrix}%
=\lambda
\begin{bmatrix}
u \\
v%
\end{bmatrix}%
,  \label{Eq4}
\end{equation}%
with $\mathcal{L}_{\text{diag}}=\frac{1}{2}\nabla _{\perp
}^{2}+V_{0}V(r,\theta )-b+2\left\vert \psi \right\vert ^{2}-3|\psi |^{4}$
and $\mathcal{L}_{\text{off}}=\psi ^{2}(1-2|\psi |^{2})$. Eigenvalues $%
\lambda $ were found from numerical solution of Eq. (\ref{Eq4}), produced by
the Fourier collocation method Yang \cite{yang20103}. Solitons are stable if
Re$(\lambda )=0$ for all eigenvalues.
\begin{figure}[tbph]
\centering
\includegraphics[width=0.42\textwidth]{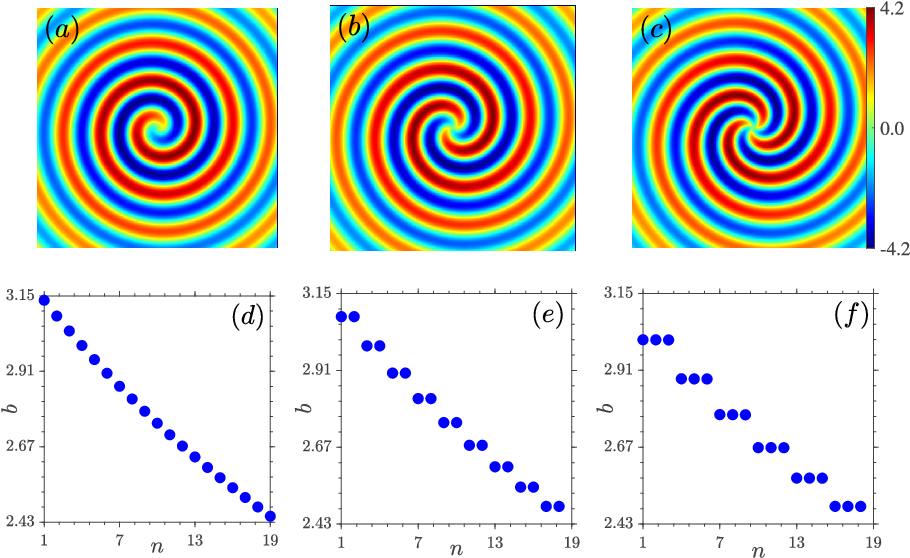} \vskip 0pc 
\caption{(a-c) The spiral potential defined by Eq. ~(\protect\ref{Eq2}).
(d-f) Spectra of discrete eigenvalues $b$ of the linearized version of Eq.~(%
\protect\ref{Eq3}). Values $n$ are sequential numbers of the eigenvalues for
$N=1$ in (a, d), $2$ in (b, e), and $3$ in (c, f). The other parameters of
the spiral potential (\protect\ref{Eq2}) are $r_{0}=6,T=2\protect\pi $, and $%
V_{0}=50$.}
\label{fig1}  \vskip -0.5pc
\end{figure}

Prior to examining the properties of nonlinear states supported by Eq.~(\ref%
{Eq1}), it is necessary to analyze its linearized version, as the respective
eigenvalues and eigenmodes provide the basis from which solitons bifurcate.
Figures~\ref{fig1}(a-c) show the spiral potentials with azimuthal indices $%
N=1,2$, and $3$.%
The potential exhibits $2N$ well-defined spiral arms whose amplitudes decay
as $r\rightarrow \infty $. Adjacent arms have opposite signs, resulting in a
structure that preserves a discrete $2N$-fold rotational anti-symmetry.


The linear spectra for the potentials from Figs.~\ref{fig1}(a-c) are shown
in Figs.~\ref{fig1}(d-f). Nonlinear fundamental states bifurcate from the first $N$ eigenmodes, while the next $N$ eigenmodes give rise to the lowest-order solitons among the higher-order families. While the eigenvalues in the potential with $N=1$ is nondegenerate [Fig.~\ref{fig1}(d)], those for $N>1$ are $N$-fold degenerate [Figs.~\ref{fig1}(e, f)].
Corresponding to each $N$-fold degenerate eigenvalue is a set of $N$
mutually orthogonal eigenmodes. Any linear superposition of the degenerate
eigenmodes also constitutes an allowed state of the linear system.
\begin{figure}[tbph]
\centering
\includegraphics[width=0.42\textwidth]{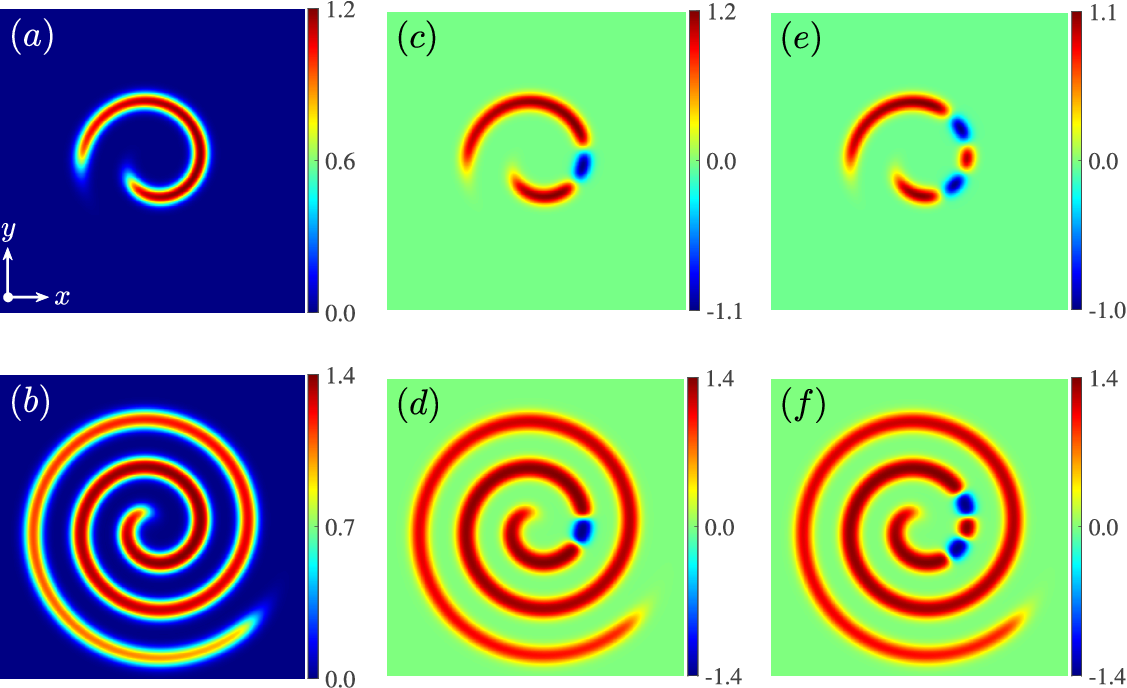} \vskip 0pc
\caption{Profiles of fundamental (a, b), third-order (c, d), and fifth-order
(e, f) spiral solitons, with $b=3.0$ in (a, c, e) and $1.9$ in (b, d, f). All
panels pertain to $N=1$ in Eq. (\protect\ref{Eq2}). Here and in other
figures,  the plotted
domain is $(x,y)\in \lbrack 20,+20]$.}
\label{fig2} \vskip -0.5pc
\end{figure}

The numerical solution of the full nonlinear equation~(\ref{Eq3}) produces
soliton families in the potential with $N=1$. Examples of fundamental
solitons (ones without internal nodes) at different values of $b$ are
displayed in Figs.~\ref{fig2}(a, b). As $b$ decreases, the soliton expands
along the potential arm, with a gradually decaying amplitude, which
indicates that the defocusing quintic nonlinearity plays a dominant role. In
our calculations, the spatial domain is chosen as $|x|,|y|\leq 20$. The
expanding soliton hits the computational boundary when $b$ falls below a
certain value. Nevertheless, the ideal soliton solution expands indefinitely
along the potential arm as $b$ decreases.

Higher-order solitons, characterized by the presence of internal nodes in
their profile along potential arms, are shown in Figs.~\ref{fig2}(c-f). They
also expand along the arm as $b$ decreases. The third-order soliton
possesses two nodes, separating field segments of opposite signs. Notably,
as $b$ decreases, the negative-valued segment between the nodes retains
nearly its original length and position, whereas the outer segments expand
substantially. A similar behavior is observed for the fifth-order soliton,
which contains four nodes and alternating signs between adjacent segments.
In both cases, the amplitude gradually decays in the outermost segments of
the solitons. The second- and fourth-order solitons, with one and three
inner nodes, respectively, were found too (not shown here).

\begin{figure}[tbph]
\centering
\includegraphics[width=0.34\textwidth]{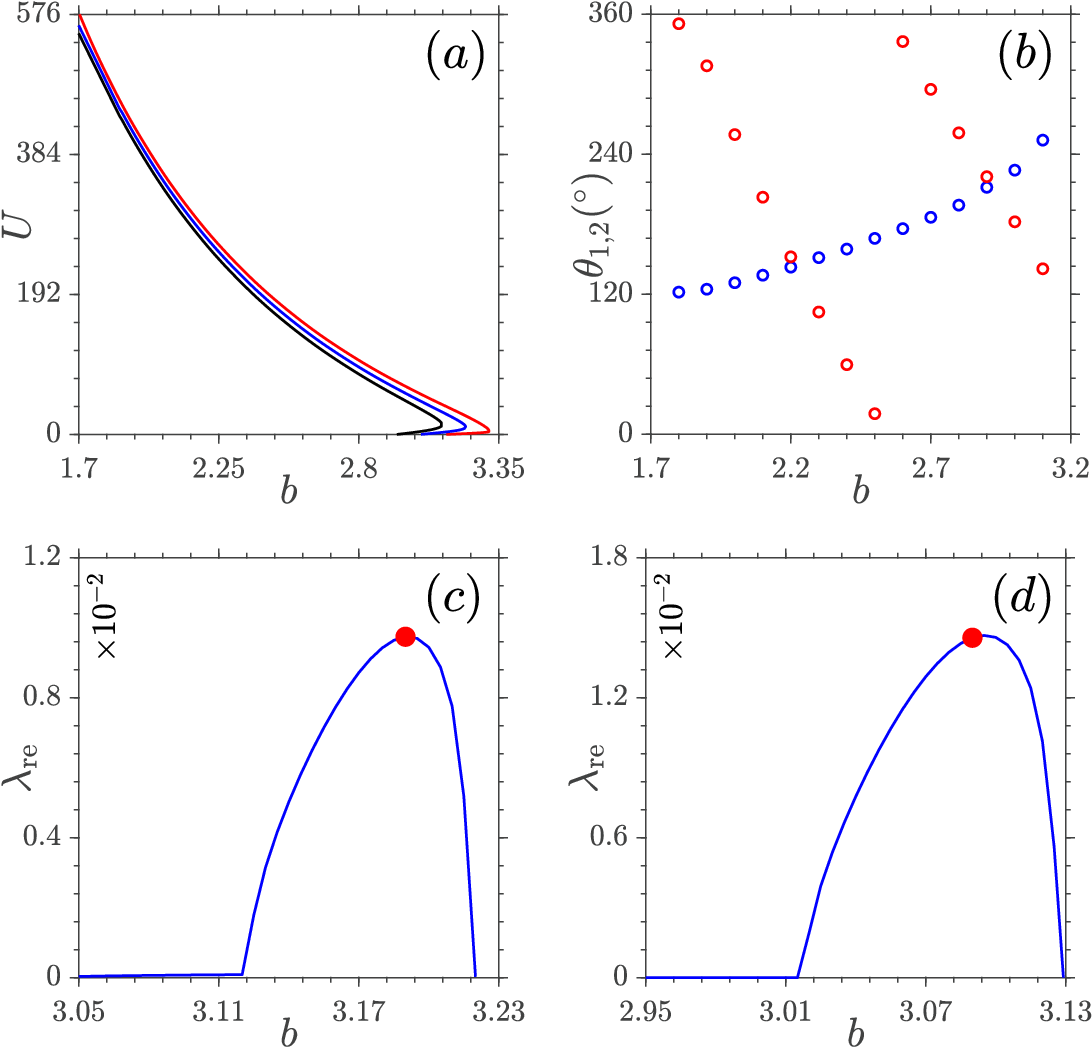} \vskip 0pc 
\caption{(a) Power $U$ vs. propagation constant $b$ for the fundamental
(red), third-order (blue), and fifth-order (black) solitons. (b) The angular
coordinates $\protect\theta _{1}$ (blue) and $\protect\theta _{2}$ (red) of
the inner and outer endpoints of the \textquotedblleft
skeleton\textquotedblright\ vs. $b$, for the upper-branch fundamental
solitons (the formal jump of $\protect\theta _{2}$ from $0^{\mathrm{o}}$ to $%
360^{\mathrm{o}}$ implies that the dependence is actually continuous). The
bottom panels: the instability growth rate vs. $b$ for the third-order (c)
and fifth-order (d) solitons belonging to the lower branch. Red dots
correspond to the propagation examples presented in Fig.~\protect\ref{fig7}.}
\label{fig3}  \vskip -0.5pc
\end{figure}

The soliton families are characterized by the $U(b)$ curves [Fig.~\ref{fig3}%
(a)]. Fundamental solitons bifurcate from the first (lowest) linear
eigenstate at $b=3.136$, while the third- and fifth-order solitons originate
from the corresponding eigenmodes at $b=3.039$ and $b=2.948$, respectively.
The bifurcation points, at which the power vanishes, precisely coincide with
the first, third, and fifth eigenvalues, respectively [see Fig.~\ref{fig1}%
(d)]. As shown by short lower branches in Fig.~\ref{fig3}(a), the power
initially increases with $b$ varying between the bifurcation point and a
turning point, at which the solitons' peak amplitude $|\psi |$$_{\text{max}}$
attains values close to $1$, hence the cubic focusing term in Eq. (\ref{Eq1}%
) is the dominant one.

Beyond the turning point, the quintic defocusing term becomes dominant in
the core region of solitons. It slows down the growth of $|\psi |$$_{\text{
max}}$ and accelerates the soliton expansion, as corroborated by Fig.~\ref%
{fig2}. The gradual transition to the defocusing, together with the
confining action of the trapping potential, prohibits the existence of
fundamental solitons for $b$ exceeding a cutoff value (which corresponds to
the turning point), i.e., at $b>b_{\text{cut}}=3.311$. The corresponding
cutoff values for third- and fifth-order solitons are $b_{\text{cut}}=3.221$
and $3.126$, respectively. Further increase of power $U$ is accommodated,
above the turning points, by upper $U(b)$ branches in Fig.~\ref{fig3}(a).

A distinctive above-mentioned feature is that solitons extend along the
spiral's arm as $b$ decreases, in contrast to solitons in radially symmetric
potentials \cite{dong2023multipole,Dong:23} or optical lattices \cite%
{Zeng2018}, where the strengthening quintic defocusing nonlinearity
typically leads to broader (fatter) soliton profiles. To quantify the
elongation of the spiral solitons, we extract a \textquotedblleft
skeleton\textquotedblright\ of the fundamental one and identify angular
positions (values of coordinate $\theta $) of its inner and outer endpoints.
The dependence of these coordinates, $\theta _{1}$ and $\theta _{2}$, on $b$
is shown in Fig.~\ref{fig3}(b). While $\theta _{2}$ varies linearly with $b$%
, $\theta _{1}$ exhibits a more complex dependence on $b$.

\begin{figure}[bh]
\centering
\includegraphics[width=0.34\textwidth]{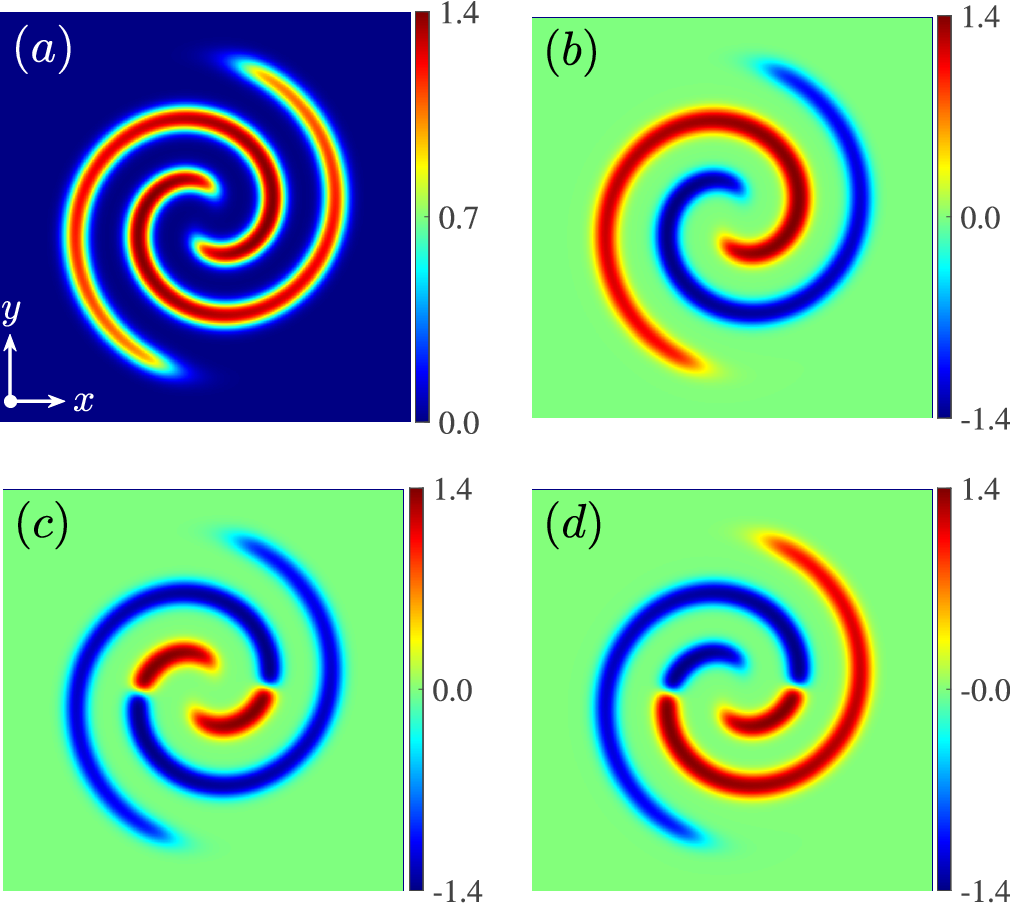} \vskip 0pc
\caption{Examples of solitons in the spiral potential (\protect\ref{Eq2})
with $N=2$. (a) and (b): The in- and out-of-phase fundamental states,
respectively. (c) and (d): The in- and out-of-phase second-order states,
respectively. All panels pertain to $b=2.2$. }
\label{fig4}  \vskip -0.5pc
\end{figure}

Our central finding is that the linear-stability analysis, based on Eqs.~(%
\ref{Eq4}), reveals that the spiral potential maintains the stability of the
fundamental and high-order solitons along the upper $U(b)$ branches, unlike
models with other potentials, where higher-order states are usually unstable
\cite{malomed2022}. Along the lower branches of the $U(b)$ dependence, the
fundamental solitons remain stable, while the higher-order ones develop a
weak instability [Figs.~\ref{fig3}(c, d)]. Note that the stability of
solitons along the lower ($dU/db>0$) and upper ($dU/db<0$) branches
qualitatively agrees with the Vakhitov-Kolokolov (VK) \cite{VK,malomed2022}
and anti-VK \cite{PhysRevA.81.013624} stability criteria, which are valid
for solitons dominated by the self-focusing and defocusing nonlinearity,
respectively.

The spiral potential $N=2$ yields a richer variety of soliton families.
Solitons can reside on a single arm or occupy two adjacent arms. The
curvature of the soliton arm is more pronounced than in the case of $N=1$.
Two-arm solitons exhibit both in-phase and out-of-phase configurations,
which feature, respectively, the same or opposite signs of $\psi \left(
x,y\right) $ in the two arms [Figs.~\ref{fig4}(a, b)]. Representative
examples of the second-order in-phase and out-of-phase solitons are
displayed in Figs.~\ref{fig4}(c) and \ref{fig4}(d), respectively. For the
in-phase solitons, the components in the two arms are identical, apart from
the difference in their spatial positions [Figs.~\ref{fig4}(a, c)], whereas
for the out-of-phase solitons the components are sign-reversed copies [Figs.~%
\ref{fig4}(b, d)]. Consequently, the in-phase solitons preserve the discrete
two-fold rotational symmetry, while the out-of-phase ones exhibit the
discrete two-fold rotational anti-symmetry.

\begin{figure}[tbph]
\centering
\includegraphics[width=0.35\textwidth]{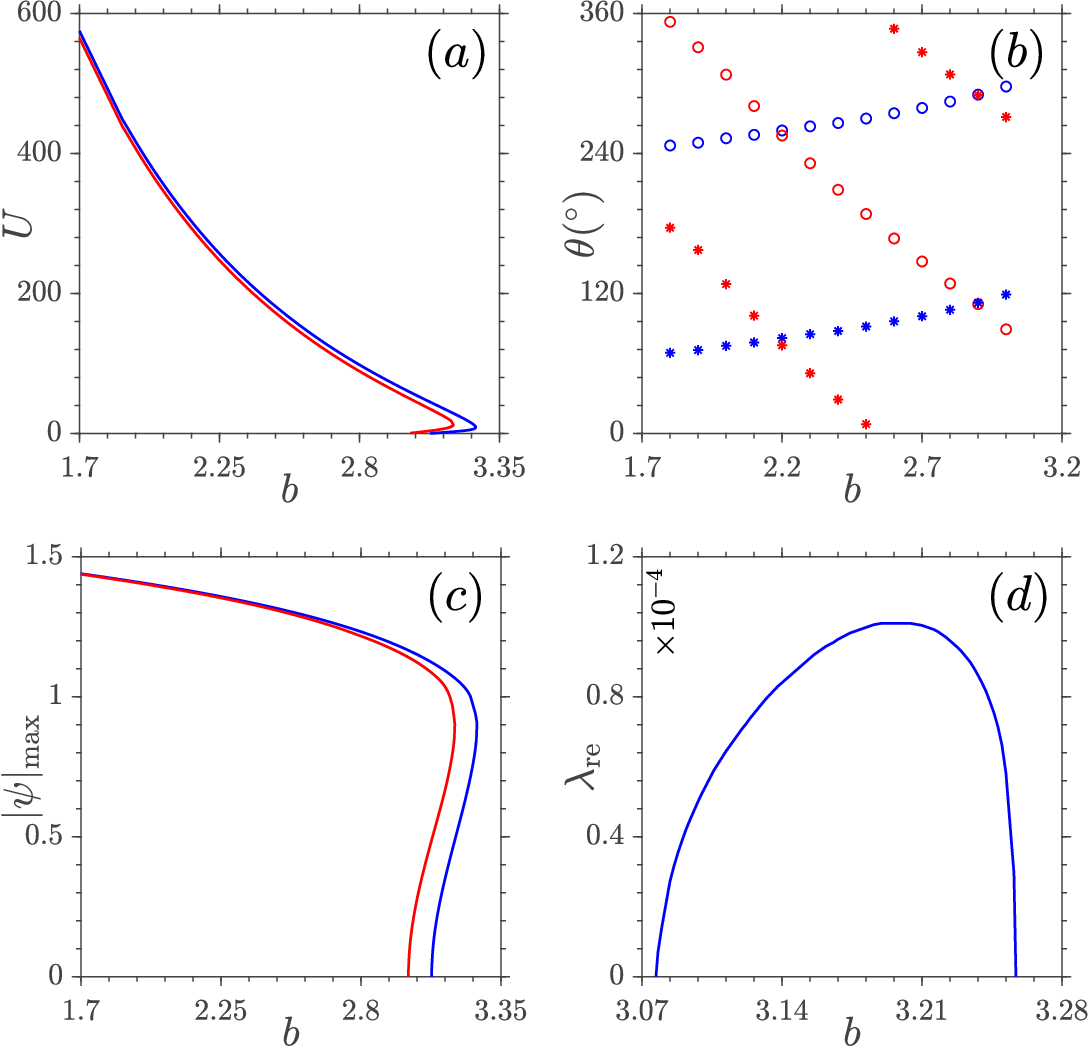} \vskip 0pc 
\caption{(a) Power $U$ vs. $b$ for the in-phase and out-of-phase fundamental
(blue) and second-order (red) two-arm ($N=2$) solitons. (b) Coordinates of the inner (blue) and outer (red) endpoints of the two arms (stars and circles, respectively) of fundamental in-phase solitons vs. $b$. (c) The peak value
$|\protect\psi |$$_{\mathrm{max}}$ of the fundamental (blue) and
second-order (red) solitons vs. $b$. (d) The instability growth rate $%
\protect\lambda _{\mathrm{re}}$ vs. $b$ for the lower-branch fundamental
in-phase solitons. }
\label{fig5} \vskip -0.5pc
\end{figure}

For two-arm solitons, identical profiles of $|\psi |$ in the in- and
out-of-phase configurations lead to the same total power dependence on $b$
[Fig.~\ref{fig5}(a)]. These solitons bifurcate from linear eigenstates at
points where power $U$ vanishes. Similar to the $N=1$ case [Fig. \ref{fig3}%
(b)], the shape of two-arm spiral solitons is determined by the angular
coordinates of the inner and outer endpoints ($\theta _{1}$ and $\theta _{2}$%
, respectively). These coordinates are plotted, as functions of $b$, in Fig.~%
\ref{fig5}(b) for the fundamental two-arm solitons. While $\theta _{2}$ is a
nearly linear function of $b$, the dependence of $\theta _{1}$ on $b$ is
more complex. Comparing the two components of the two-arm solitons, we
conclude that differences between their inner and outer endpoints are $%
\Delta \theta _{1}=\Delta \theta _{2}=\pi $ (see Fig. \ref{fig4}), which
manifests the $\pi $-rotation symmetry of $N=2$ potential.

The turning points in Fig.~\ref{fig5}(c) are determined by the competition
between the cubic and quintic terms in Eq. (\ref{Eq3}), therefore $|\psi |$$%
_{\mathrm{max}}\approx 1$ at these points. The gradual expansion of the
solitons along the arms prevents the rapid growth of the peak amplitude
[Fig.~\ref{fig5}(c)], thereby favoring the stability of the solitons. The
linear-stability analysis confirms that all upper-branch solitons
(fundamental and higher-order ones alike) possess zero instability growth
rates, $\lambda _{\mathrm{re}}=0$. Lower-branch solitons display very weak
instability, with $\lambda _{\mathrm{re}}\sim 10^{-4}$ [Fig.~\ref{fig5}](d),
allowing the weakly unstable soliton to persist in the course of the
propagation.
\begin{figure}[tbph]
\centering
\includegraphics[width=0.4\textwidth]{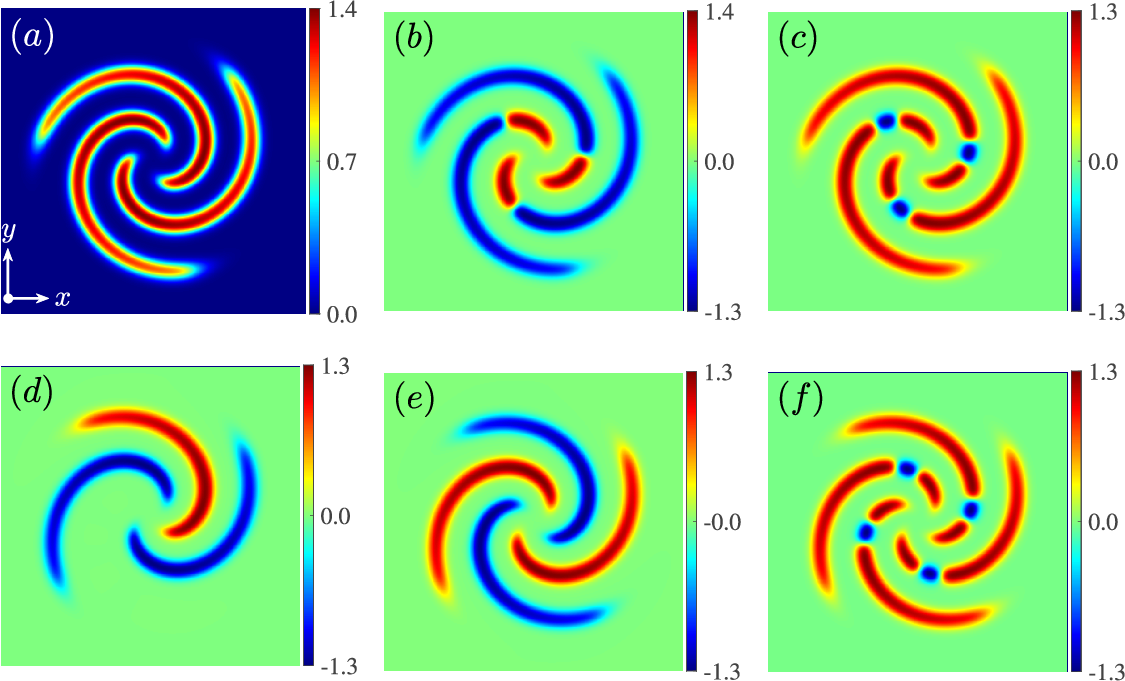} \vskip 0pc 
\caption{Profiles of solitons supported by the spiral potential (\protect\ref%
{Eq2}) with $N=3$ (a-c) and $4$ (d-f). (a) A fundamental in-phase state. (b)
A second-order in-phase state. (c) A third-order in-phase state. (d) A
hybrid-phase state, with one unpopulated arm. (e) An out-of-phase state. (f)
A third-order\ in-phase state. All panels pertain to $b=2.2$. }
\label{fig6}  \vskip -0.8pc
\end{figure}

The further increase of $N$ in Eq. (\ref{Eq2}) leads to a still broader
variety of soliton configurations [Fig. \ref{fig6}]. For $N=3$, besides the
in-phase fundamental and higher-order solitons [Figs.~\ref{fig6}(a-c)],
hybrid-phase modes, built of two in-phase components and one with the
opposite sign, exist as well. For $N=4$, an example is displayed in Fig.~\ref%
{fig6}(d), where one of the four potential arms remains unpopulated.
Fundamental out-of-phase and higher-order in-phase solitons are displayed in
Figs.~\ref{fig6}(e) and (f), respectively. With the decrease of $b$, all
families exhibit expansion along the potential arms. The comprehensive
linear-stability analysis verifies that all species of upper-branch solitons
with $N\geq 3$ remain entirely stable in their existence regions.

\begin{figure}[tbph]
\centering
\includegraphics[width=0.38\textwidth]{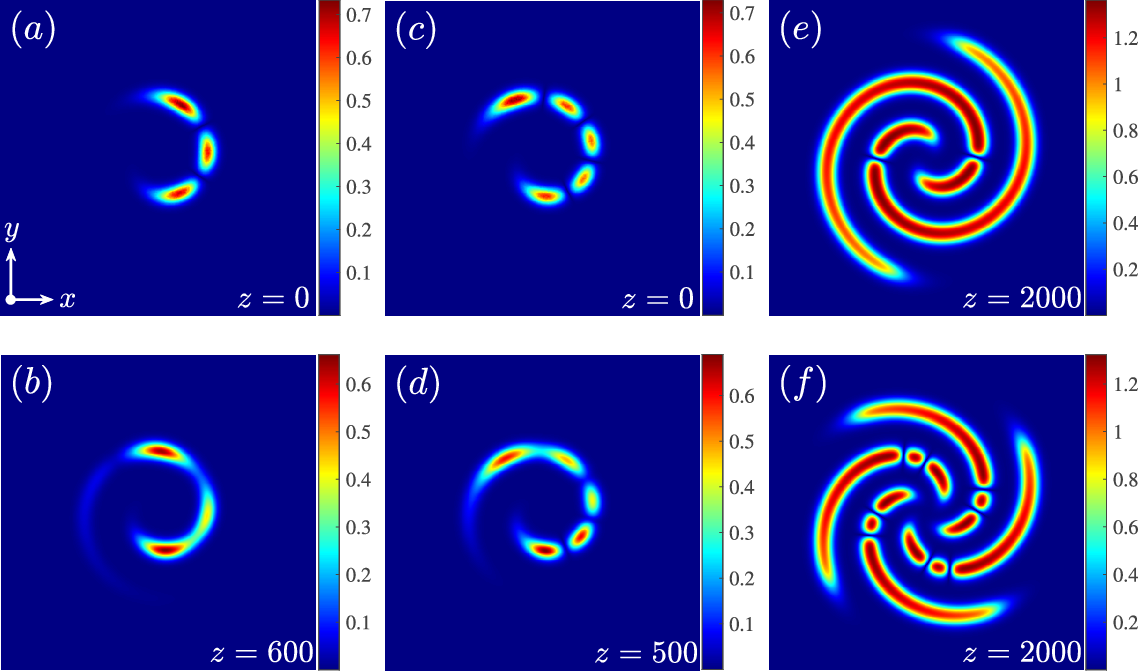} \vskip 0pc 
\caption{(a-d): The perturbed evolution of unstable lower-branch third-order
(a,b) and fifth-order (c,d) solitons, with $N=1$, marked in Figs.~\protect
\ref{fig3}(c, d). (e, f): The results of the stable propagation at $z=2000$
for the upper-branch\ second- and third-order solitons shown in Figs.~%
\protect\ref{fig4}(b) ($N=2$) and \protect\ref{fig6}(f) ($N=4$),
respectively. $b=3.19$ in (a, b), $3.09$ in (c, d), and $2.2$ in (e, f).}
\label{fig7} \vskip -0.5pc
\end{figure}

To verify the predictions of the linear-stability analysis, extensive
simulation of perturbed propagation of the solitons were conducted by means
of the split-step Fourier method. Typical examples of unstable lower-branch
and stable upper-branch solitons are presented in Fig.~\ref{fig7}. For the
lower-branch third- and fifth-order solitons in the potential (\ref{Eq2})
with $N=1$, weak instability with growth rates $\lambda _{\mathrm{re}}=0.0097
$ and $0.0146$ destroy the original modes from Figs.~\ref{fig7}(a) and (c),
after a relatively long propagation, as shown in Figs.~\ref{fig7}(b) and
(d), respectively. On the other hand, stable upper-branch higher-order
solitons maintain their shapes intact after arbitrarily long propagation
[Figs.~\ref{fig7}(e, f)]. Thus, the predictions of the linear-stability
analysis have been completely corroborated by the direct simulations in all
considered cases.

Summarizing, we have studied the structure and propagation of solitons in
the cubic-quintic medium equipped with the spiral trapping potential. The
increase of the number of spiral's arms $N$ extends the variety of the
solitons. With the increase of power, the solitons expand along the
potential's arms. All upper-branch solitons are stable in their entire
existence domains. The present analysis can be extended to matter-wave
solitons and quantum droplets in atomic BEC loaded into a spiral potential.
A challenging prospect is to generalize the study to spatiotemporal solitons
supported by spiral potentials.

\vskip0.2pc \noindent\textbf{Funding.} Natural Science Foundation of China
(NSFC) (62575264); Applied Basic Research Program of Shanxi Province
(202303021211191). %

\vskip0.2pc \noindent \textbf{Disclosures.} The authors declare no conflicts of interest.

\newpage \noindent\textbf{{\Large {References with titles}}}

\end{document}